\documentclass[12pt]{iopart}
\usepackage{iopams}
\usepackage{graphicx}
\begin{document}
\jl{3}

\title[Superconductivity and spin-glass like behavior in Pd-MG]
{Superconductivity and spin-glass like behavior in system with 
Pd sheet sandwiched between graphene sheets}

\author{Masatsugu Suzuki\dag\footnote[3]{suzuki@binghamton.edu}, Itsuko S.
Suzuki\dag  and J\"{u}rgen
Walter\ddag\footnote[4]{Juerg\_Walter@t-online.de}}

\address{\dag\ Department of Physics, State University of New York at Binghamton,
Binghamton, New York 13902-6016, USA}

\address{\ddag\ Department of Materials Science and Processing, Graduate School 
of Engineering, Osaka University, 2-1, Yamada-oka, Suita, 565-0879, JAPAN}

\begin{abstract}
Pd-metal graphite (Pd-MG) has a layered structure, where each Pd 
sheet is sandwiched between adjacent graphene sheets. 
DC magnetization and AC magnetic susceptibility of Pd-MG
have been measured using a SQUID magnetometer. 
Pd-MG undergoes a superconducting transition at $T_{c}$ ($= 3.63 \pm 0.04$
K).  The superconductivity occurs in Pd sheets.  The relaxation of
$M_{ZFC}$ (aging), which is common to spin glass systems, is also observed
below $T_{c}$.  The relaxation rate $S(t)$ shows a peak at a characteristic
time $t_{cr}$, which is longer than a wait time $t_{w}$.  The
irreversibility between $\chi_{ZFC}$ and $\chi_{FC}$ occurs well above
$T_{c}$.  The susceptibility $\chi_{FC}$ obeys a Curie-Weiss behavior with
a negative Curie-Weiss temperature ($-13.1 \leq \Theta \leq -5.4$ K).  The
growth of antiferromagnetic order is limited by the disordered nature of
nanographites, forming spin glass-like behavior at low temperatures in
graphene sheets.
\end{abstract}



\maketitle

\section{\label{intro}Introduction}
Magnetism and superconductivity are manifestations of two different ordered
states into which metals can condense at low temperatures.  In general
these two states are mutually exclusive; they do not coexist at the same
site in the system.  The study of the interplay between these properties
has recently been revitalized by the discovery of a family of high-$T_{c}$
boride carbides RNi$_{2}$B$_{2}$C (where R is a rare-earth
element) \cite{Nagarajan1994,Cava1994}. The R-C layers separated by
Ni$_{2}$B$_{2}$ sheets are antiferromagnetically ordered below a N\'{e}el
temperature $T_{N}$, while Ni$_{2}$B$_{2}$ sheets are superconducting below
$T_{c}$.  Like RNi$_{2}$B$_{2}$C, Pd-metal graphite (MG) has a layered
structure \cite{Walter1999,Walter2000a,Walter2000b,Walter2000c,Suzuki2000}.
Ideally, Pd monolayer is sandwiched between adjacent graphite sheets,
forming a periodic $c$-axis stacking like graphite intercalation compounds
(GIC's).  Each Pd layer consists of small islands formed of Pd
nanoparticles with finite sizes.  Pd nanoparticles would generate internal
stress inside the graphite lattice, leading to the break up of adjacent
graphene sheets into nanographites.  
It is expected that the superconductivity occurs in Pd
sheets and that the antiferromagnetic (AF) short-range order occurs in
nanographites in graphene sheets, leading to a spin glass-like behavior.

In the present work, we have undertaken an extensive study on the magnetic
properties of Pd-MG from various kinds of measurement using SQUID
magnetometer: the DC magnetization in the zero-field cooled (ZFC),
field-cooled (FC), thermoremnant (TR), and isothermal remnant (IR) states,
and the AC magnetic susceptibility (the dispersion and absorption).  We
show that this compound undergoes a superconducting transition at a
critical temperature $T_{c}$ ($= 3.63 \pm 0.04$ K).  The superconductivity
occurs in Pd sheets.  The relaxation of ZFC magnetization $M_{ZFC}(t)$
is also observed below $T_{c}$.  It depends on a wait time $t_{w}$,
a time spent at constant temperature before the magnetic field is applied. 
This phenomenon is called aging and has been observed in spin glass (SG)
systems.  As far as we know, the aging effect has not been observed in
usual superconductors.  There is only one exception for a certain
Bi$_{2}$Sr$_{2}$CaCu$_{2}$O$_{8}$ sample displaying a paramagnetic Meissner
effect \cite{Papadopoulou1998,Papadopoulou2001}. In this sense, the aging
behavior in Pd-MG is considered to be magnetic but not superconducting in
origin.  The FC susceptibility well above $T_{c}$ obeys a Curie-Weiss
behavior with a negative Curie-Weiss temperature $\Theta$.  The deviation
of the ZFC susceptibility from the FC susceptibility starts to occur even
at 298 K. These results suggest that an AF short-range order appears well
above $T_{c}$.  Based on these results, we propose a model that the
superconductivity occurs in Pd sheets and that the AF short-range order
occurs in graphene sheets.  Although no long-range magnetic order below
$T_{c}$ has been clearly confirmed from the present work, it is assumed
that the possible interplay between the superconductivity and the SG-like
behavior becomes significant below $T_{c}$.

Here we present experimental and theoretical backgrounds for the origin of
the superconductivity in Pd sheets and the antiferromagnetism in
nanographites.  The absence of superconductivity in pristine Pd above 2 mK
is mainly due to strong spin fluctuations \cite{Fay1977}. Theoretically it
is suggested that Pd without spin fluctuations should be a
superconductor \cite{Papaconstantopoulos1975,Pinski1978}. Experimentally,
Stritzker \cite{Stritzker1979} has reported that pure Pd films,
evaporated between 4.2 and 300 K, can be transformed into superconductors
by means of irradiation at low temperatures with He$^{+}$ ions.  The
maximum transition temperature obtained is 3.2 K. A special kind of
disorder produced by low temperature irradiation may lead to a smearing of
the Fermi energy $E_{F}$, and thus to a reduction of the density of states
(DOS) at $E_{F}$, $N(E_{F})$.  This reduction of $N(E_{F})$ leads to a
decrease in the Stoner enhancement factor.  As a result, the strong spin
fluctuations would be reduced and superconductivity might be possible.  In
fact, Meyer and Stritzker \cite{Meyer1982} have shown that the AC
magnetic susceptibility of low-temperature irradiated Pd is strongly
reduced in comparison to the annealed Pd metal.

Nanographites are nanometer-sized graphite fragments, forming a new class
of mesoscopic system.  Fujita et al. \cite{Fujita1996} and Wakabayashi et
al. \cite{Wakabayashi1999} have theoretically suggested that the electronic
structures of finite-size graphene sheets depend crucially on the shape of
their edges.  Finite graphite systems having zigzag edges exhibit a special
edge state.  The corresponding energy bands are almost flat at $E_{F}$,
giving a sharp peak in $N(E_{F})$.  The conduction electrons
localized near the zigzag edge have magnetic moments (= $g\mu_{B}S$ with $g
= 2$ and $S = 1/2$). 
Harigaya \cite{Harigaya2001a,Harigaya2002,Harigaya2001b} has theoretically
predicted that the magnetism in nanographites with zigzag edge sites
depends on the stacking sequence of nanographites.  The structure of
pristine graphite consists of hexagonal net planes of carbon stacked along
the $c$ axis in a staggered array usually denoted as
$ABAB\dots$ \cite{Enoki2003}. There is a lateral shift on going from layer
$A$ to layer $B$.  There are two kinds of spin configurations depending on
the stacking of nanographites.  For the $A$-$B$ stacking, there is no
interlayer interaction at the edge site, giving rise to the AF spin
alignment.  The growth of AF spin order is greatly limited by the
disordered nature of nanographites, forming AF short-range order.  For the
$A$-$A$ type stacking, on the other hand, the magnetic moment per layer
does not appear due to the interlayer interaction.

\section{\label{exp}EXPERIMENTAL PROCEDURE}
In a previous work \cite{Suzuki2000} we have shown that there are two types
of Pd-MG depending on the reduction condition in the sample preparation:
the long-reaction time and the short-reaction time.  The physical
properties of these two systems are rather different: the ferromagnetic
nature for the long-reaction time and the AF nature for the short-reaction
time.  The sample used in the present work (the short-reaction time) is the
same as that used in the previous work \cite{Suzuki2000}.  Pd-MG samples
based on natural graphite were prepared by heating PdCl$_{2}$ GIC with
mixed stages (2, 3, and 4) at 350 $^\circ$C under hydrogen gas (flow rate
300 ml per minute) for 2 hours.

The detail of sample characterization for Pd-MG was presented in the
previous papers \cite{Walter1999,Walter2000b}.  The average size of Pd
nanoparticles was estimated as ($530 \pm 340$) {\AA} from the bright field
transmission electron microscope photograph \cite{Walter1999}.  The
existence of nanographites was confirmed from the Raman scattering
\cite{Walter2000b}.  The Raman spectrum shows a large peak at 1580
cm$^{-1}$ assigned as the $E_{2g2}$ mode in pristine graphite (1582
cm$^{-1}$), and a small peak at 1360 cm$^{-1}$ assigned as the D-band
(disordered induced modes) of graphite \cite{Walter2000b}.  The size of
nanographites can be estimated from an empirical law $L_{a} = 4.4 \times
I_{1580}/I_{1350}$, where $I_{1580}$ and $I_{1360}$ are the intensities of
the corresponding peaks.  In fact, the size of nanographites is on the same
order as that of Pd nanoparticles.  There is no appreciable charge transfer
between nanographites and Pd nanoparticles.

The Pd-MG sample consists of many small flakes.  Each flake has a
well-defined $c$ axis.  If these flakes are carefully piled inside the
sample capsule for the measurement, the $c$ axis of the whole sample could
correspond to the $c$ axis of each flake.  This is not the case for the
present experiment.  The present sample may be regarded as a powdered
sample with the $c$ axis randomly distributed over all directions.  Because
of the small samples, the resistivity measurement could not be carried out. 
The mass of the sample used in the present work was 23.8 mg.

The DC magnetization and AC magnetic susceptibility of Pd-MG were measured
using a SQUID magnetometer (Quantum Design, MPMS XL-5).  Before setting up
a sample at 298 K, a remnant magnetic field was reduced to less than 3 mOe
using an ultra-low-field capability option.  For convenience, hereafter
this remnant field is denoted as the state $H$ = 0.  (i) \textit{DC
magnetization}.  The sample was cooled from 298 to 1.9 K at $H$ = 0.  After
an external magnetic field ($H$) was applied at 1.9 K, the zero-field
cooled magnetization ($M_{ZFC}$) was measured with increasing $T$ from 1.9
to 60 K. The sample was kept at 70 K for 20 minutes.  Then the field cooled
magnetization ($M_{FC}$) was measured with decreasing $T$ from 60 to 1.9 K.
(ii) \textit{A hysteresis loop of DC magnetization}.  The sample was cooled
from 298 K to $T$ (= 1.9 or 3.3 K) at $H$ = 0.  Then DC magnetization at
$T$ was measured as $H$ was varied from $H$ = 0 to 1 kOe, from 1 to -1 kOe,
and from -1 to 1 kOe in order.  (iv) \textit{AC magnetic
susceptibility} ($\chi = \chi^{\prime} + i\chi^{\prime\prime}$).  The
sample was cooled from 298 to 1.9 K at $H = 0$.  Then the dispersion
$\chi^{\prime}$ and absorption $\chi^{\prime\prime}$ were simultaneously
measured with increasing $T$ from 1.9 to 20 K with and without $H$, where
the frequency and amplitude of the AC magnetic field were $f = 1$ Hz and $h
= 2$ Oe, respectively.  After each $T$ scan, $H$ was changed at 30 K. The
sample was cooled from 30 to 1.9 K. Then the measurement was repeated with
increasing $T$ from 1.9 to 20 K in the presence of $H$.

\section{\label{result}RESULT}
\subsection{\label{resultA}$\chi^{\prime}$ and $\chi^{\prime\prime}$}

\begin{figure}
\begin{center}
\includegraphics[width=8.0cm]{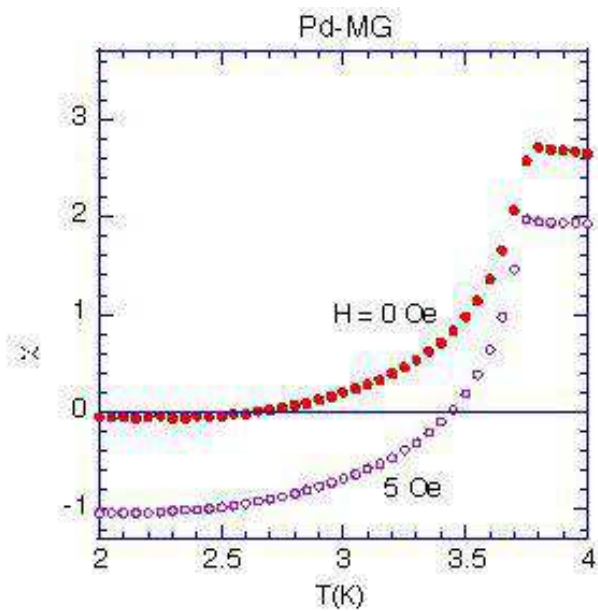}%
\end{center}
\caption{\label{fig:one}$T$ dependence of the dispersion $\chi^{\prime}$
for Pd-MG at $H = 0$ and 5 Oe.  $f = 1$ Hz.  $h = 2$ Oe.}
\end{figure}

\begin{figure}
\begin{center}
\includegraphics[width=12.0cm]{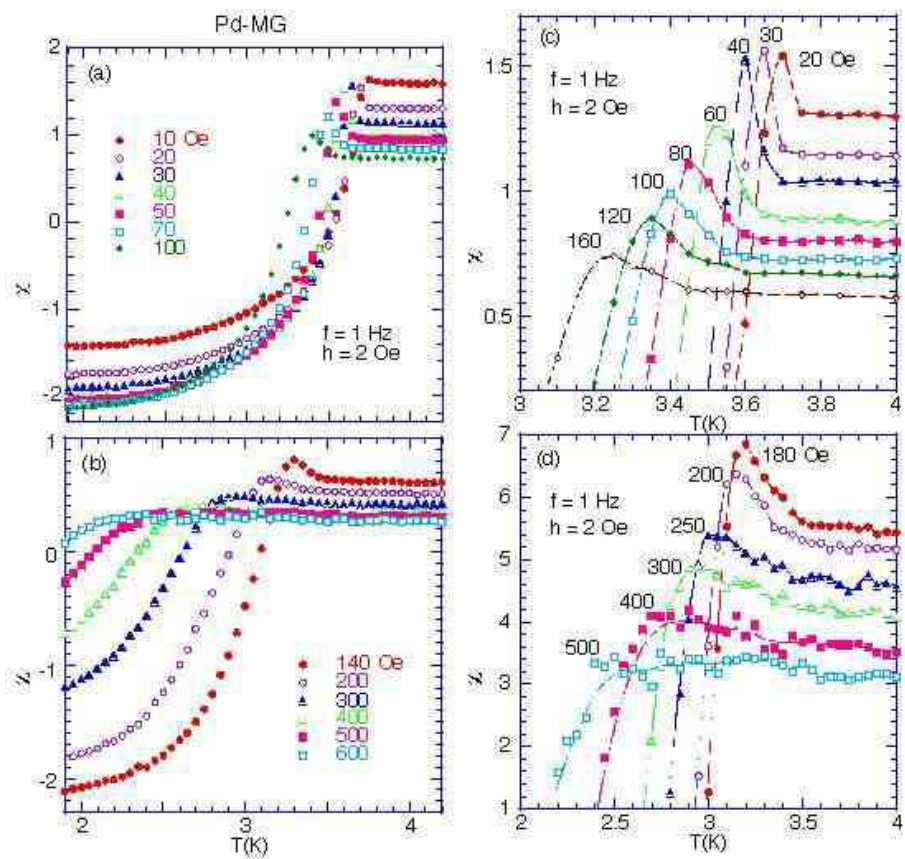}%
\end{center}
\caption{\label{fig:two}(a) - (d) $T$ dependence of $\chi^{\prime}$ for
Pd-MG at various $H$.  $f = 1$ Hz.  $h = 2$ Oe.  The solid lines in (c) and
(d) are guides to the eyes.}
\end{figure}

Figures \ref{fig:one} and \ref{fig:two} show the $T$ dependence of
$\chi^{\prime}$ for Pd-MG at various $H$, where $f = 1$ Hz and $h = 2$ Oe. 
The dispersion $\chi^{\prime}$ at $H = 0$ slightly increases with
decreasing $T$ at the high-$T$ side.  It shows a peak at 3.8 K. Below 3.8
K, $\chi^{\prime}$ drastically decreases with further decreasing $T$.  The
sign of $\chi^{\prime}$ changes from positive to negative below 2.65 K. In
contrast, the dispersion $\chi^{\prime}$ for $5 \leq H \leq 350$ Oe still
shows a positive peak, shifting to the low-$T$ side with increasing $H$. 
However, the sign of $\chi^{\prime}$ changes from positive to negative
below a zero-crossing temperature $T_{0}$ (= 3.45 K at $H$ = 5 Oe).  For
$400\leq H\leq 600$ Oe, no appreciable peak is observed in $\chi^{\prime}$. 
The dispersion $\chi^{\prime}$ starts to decrease with decreasing $T$ below
$T_{0}$.  The negative sign of $\chi^{\prime}$ below $T_{0}$ for $H\geq 5$
Oe is related to a diamagnetic flux expulsion (the Meissner effect), giving
a bit of evidence of the superconductivity at low temperatures.  We find
that the derivative d$\chi^{\prime}$/d$T$ shows a peak, shifting to the
low-$T$ side with increasing $H$.  The
peak temperature of d$\chi^{\prime}$/d$T$ vs $T$ is regarded as a
superconducting transition temperature $T_{c}(H)$ ($= T_{2}(H)$) (see
section \ref{disA} for the definition of $T_{2}(H)$).  Using the value of
$\chi^{\prime}$ at 1.9 K ($\approx - 2 \times 10^{-5}$ emu/g) and the
density $\rho$ which is on the order of 1- 2 g/cm$^{3}$,\cite{Suzuki2000}
the fraction of flux expulsion relative to complete diamagnetism ($\chi_{0}
= -1/4\pi = -0.0796$ emu/cm$^{3}$) is estimated as only 0.05 \%, suggesting
isolated superconducting islands.

\begin{figure}
\begin{center}
\includegraphics[width=8.0cm]{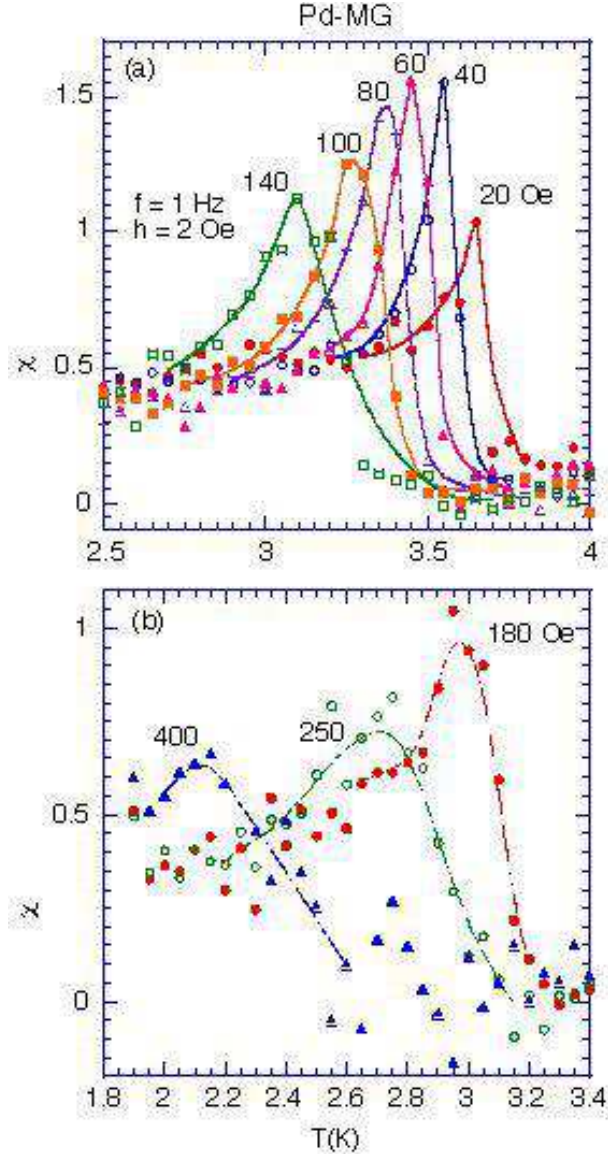}%
\end{center}
\caption{\label{fig:three}(a) and (b) $T$ dependence of the absorption
$\chi^{\prime\prime}$ for Pd-MG at various $H$.  $f = 1$ Hz.  $h = 2$ Oe. 
The solid lines are guides to the eyes.}
\end{figure}

Figure \ref{fig:three} shows the $T$ dependence of the absorption
$\chi^{\prime\prime}$ in the presence of $H$.  The data at $H = 0$ were
presented in the previous paper \cite{Suzuki2000}.  The absorption
$\chi^{\prime\prime}$ at $H$ = 0 increases with decreasing $T$.  A drastic
increase is observed around 3.5 K. In contrast, the $T$ dependence of
$\chi^{\prime\prime}$ at $H \geq 20$ Oe is rather different from that at $H
= 0$.  It has a relatively sharp peak at a peak temperature.  This peak
shifts to the low-$T$ side with increasing $H$.  The peak temperature is
the same as $T_{2}(H)$.  We notice that the peak
temperatures of $\chi^{\prime}$ and $\chi^{\prime\prime}$ at $H = 0$ are
almost independent of AC frequency $f$ for 0.07 Hz $\leq f \leq 1$ kHz as
have been reported in the previous paper \cite{Suzuki2000}.

\subsection{\label{resultB}$M$-$H$ curve}

\begin{figure}
\begin{center}
\includegraphics[width=8.0cm]{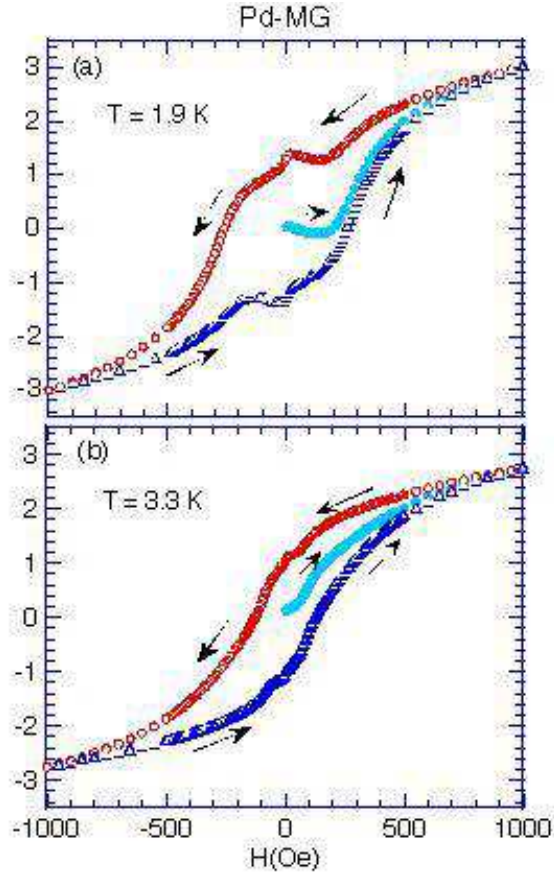}%
\end{center}
\caption{\label{fig:four}Hysteresis loop of DC magnetization $M$ for Pd-MG
at (a) $T = 1.9$ K and (b) 3.3 K. After the sample was cooled from 298 to
1.9 K at $H = 0$, the measurement was made with increasing $H$ from 0 to 1
kOe (closed circles), with decreasing $H$ from 1 to -1 kOe (open circles),
and with increasing $H$ from -1 to 1 kOe (open triangles).}
\end{figure}

\begin{figure}
\begin{center}
\includegraphics[width=8.0cm]{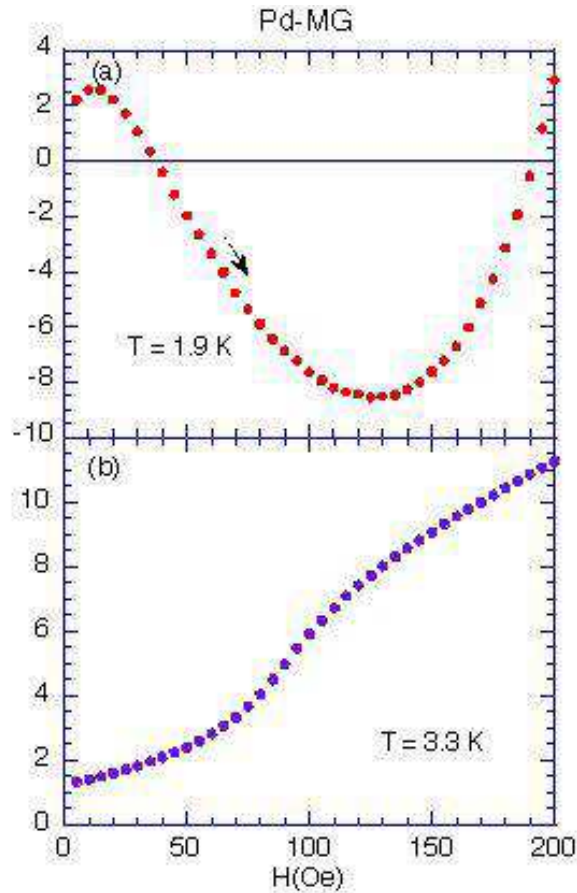}%
\end{center}
\caption{\label{fig:five}$H$ dependence of DC magnetization $M_{ZFC}$ for
Pd-MG at (a) 1.9 K and (b) 3.3 K (closed circles).  After the sample was
cooled from 298 to 1.9 K at $H = 0$, the measurement was carried out with
increasing $H$ from 0 to 200 Oe.  The data around $H = 0$ at 1.9 K is
slightly different from that shown in figure \ref{fig:four}.}
\end{figure}

Figure \ref{fig:four} shows the hysteresis loop of the DC magnetization
$M(H)$ at $T$ = 1.9 and 3.3 K, respectively.  After the sample was cooled
from 298 to 1.9 K at $H$ = 0, the magnetization $M(H)$ was measured for $-1
\leq H \leq 1$ kOe.  The magnetization curve at 1.9 K shows a non-typical
hysteresis behavior.  There are some anomalies at low-field region in the
demagnetization and remagnetization curves at 1.9 and 3.3 K, which
completely disappear at 5.5 K. In figures \ref{fig:five}(a) and (b) we show
the detail of the initial magnetization curve at 1.9 and 3.3 K. For the
data at $T$ = 1.9 K, the positive value of the initial magnetization
$M_{int}$ around $H$ = 15 Oe is sensitive to the condition of cooling from
298 to 1.9 K at $H = 0$.  The negative value of $M_{int}$ for $38 \leq H
\leq 190$ Oe gives a bit of evidence for the occurrence of the Meissner
effect.  The local-minimum field may be related to the lower critical field
$H_{c1}$.  At 3.3 K this negative local minimum disappears.

\subsection{\label{resultC}IRM and TRM}

\begin{figure}
\begin{center}
\includegraphics[width=12.0cm]{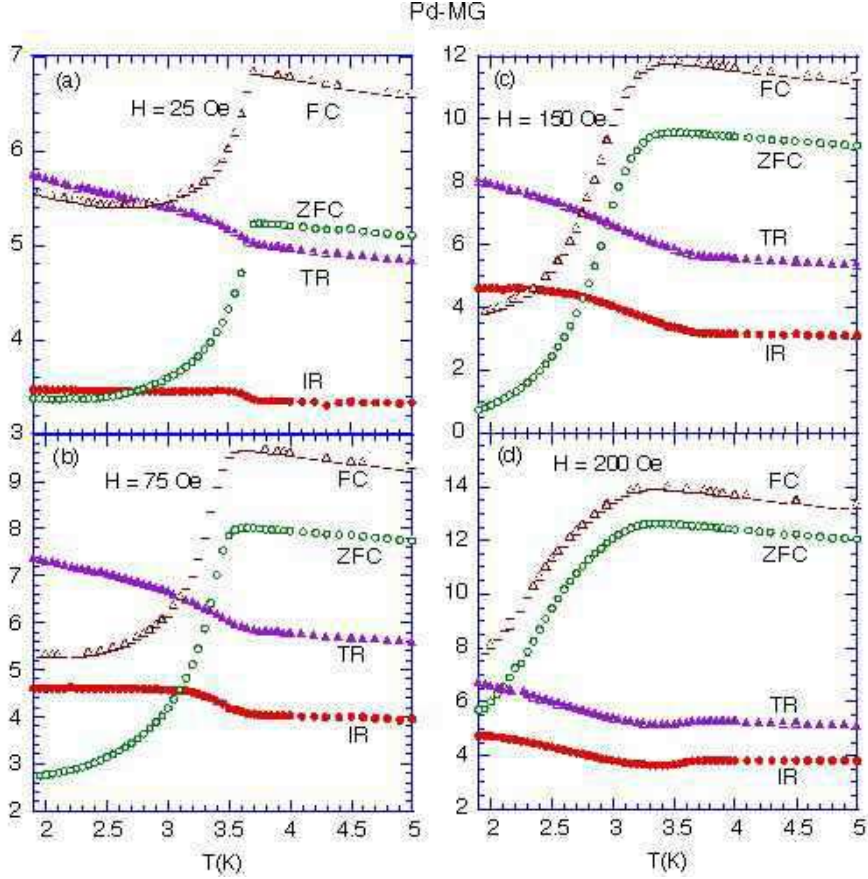}%
\end{center}
\caption{\label{fig:six}$T$ dependence of $M$$_{ZFC}$, $M$$_{IR}$,
$M$$_{FC}$, and $M$$_{TR}$ for Pd-MG. (a) $H$ = 25, (b) 75, (c) 150, and
(d) 200 Oe.  During the ZFC measurement, $M$$_{ZFC}$ at each $T$ was
measured at $H$ and $M$$_{IR}$ was measured 10$^{2}$ sec later after $H$
was set to zero field.  During the FC measurement, $M$$_{FC}$ at each $T$
was measured at $H$.  $M$$_{TR}$ was measured 10$^{2}$ sec later after $H$
was set to zero field.}
\end{figure}

Figure \ref{fig:six} shows the $T$ dependence of the magnetization
$M_{ZFC}$, $M_{IR}$, $M_{FC}$, and $M_{TR}$ for $H$ = 25, 75, 150, and 200
Oe.  The magnetization $M_{IR}$ is the isothermal remnant (IR)
magnetization.  During the process of ZFC measurement (with increasing
$T$), $M_{ZFC}$ at each $T$ was measured at $H$ and $M_{IR}$ was measured
at the same $T$ 10$^{2}$ sec later after setting $H$ to zero.  The
magnetization $M_{TR}$ is the thermoremnant (TR) magnetization.  During the
process of FC measurement (with decreasing $T$), $M_{FC}$ at each $T$ was
measured at $H$ and $M_{TR}$ was measured at the same $T$ 10$^{2}$ sec
later after setting $H$ to zero.  The magnetization $M_{TR}$ is larger than
$M_{FC}$ and $M_{IR}$ is larger than $M_{ZFC}$ below a characteristic
temperature $T_{\alpha}$ ($T_{\alpha}$ = 3.1 K for $H$ = 75 Oe), which are
features common to the superconducting state.  Similar behavior of $M_{TR}$
vs $T$ and $M_{FC}$ vs $T$ has been reported in high $T_{c}$
superconductors La$_{2}$CuO$_{4-y}$:Ba \cite{Muller1987} and
Bi$_{2}$Sr$_{2}$CaCu$_{2}$O$_{8}$ \cite{Papadopoulou1998,Papadopoulou2001}. 
This is due to the flux trapping after switching off the field, giving a
strong evidence of the superconductivity at low temperatures.  The value of
$T_{\alpha}$ is a little lower than $T_{2}(H)$: $T_{2}(H)$ = 3.34 K at $H$
= 75 Oe.  Above $T_{\alpha}$, $M_{FC}$ is larger than $M_{TR}$ and
$M_{ZFC}$ is larger than $M_{IR}$, which are features common to spin
systems with spin frustration effects.  The derivatives d$M_{TR}$/d$T$ and
d$M_{IR}$/d$T$ at $H$ = 75 Oe exhibit negative local minima at 3.34 K and
3.45 K, respectively.  The magnetization $M_{TR}$ decreases with increasing
$T$ and reduces to a positive value above $T_{2}(H)$, suggesting the
existence of AF short-range order.  This is in contrast to the case of
La$_{2}$CuO$_{4-y}$:Ba \cite{Muller1987} and
Bi$_{2}$Sr$_{2}$CaCu$_{2}$O$_{8}$ \cite{Papadopoulou1998,Papadopoulou2001}:
$M_{TR}$ becomes zero above $T_{c}(H)$ (= $T_{2}(H)$), since no flux
trapping occurs.

\subsection{\label{resultD}$\chi_{ZFC}$ and $\chi_{FC}$}

\begin{figure}
\begin{center}
\includegraphics[width=8.0cm]{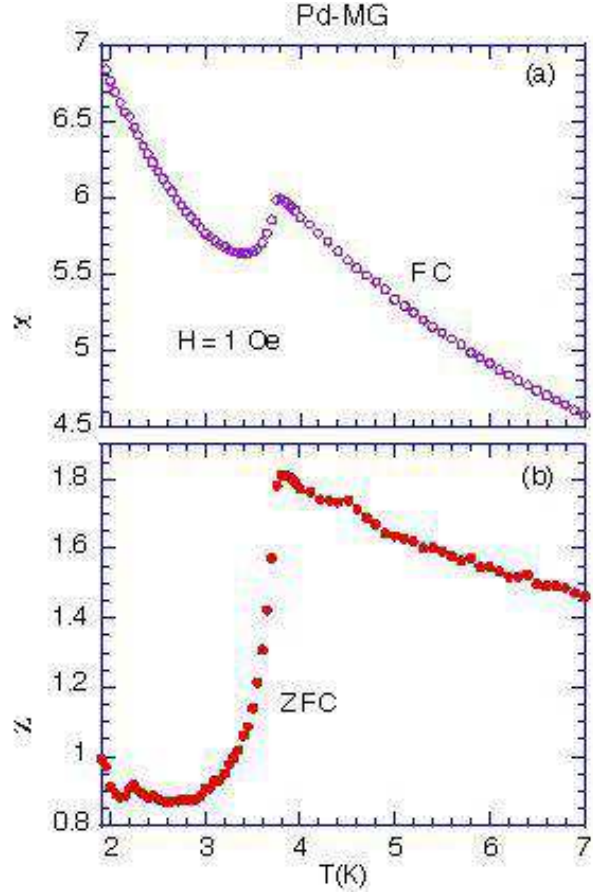}%
\end{center}
\caption{\label{fig:seven}$T$ dependence of (a) $\chi_{ZFC}$ and (b)
$\chi_{FC}$ for Pd-MG. $H$ = 1 Oe.}
\end{figure}

\begin{figure}
\begin{center}
\includegraphics[width=12.0cm]{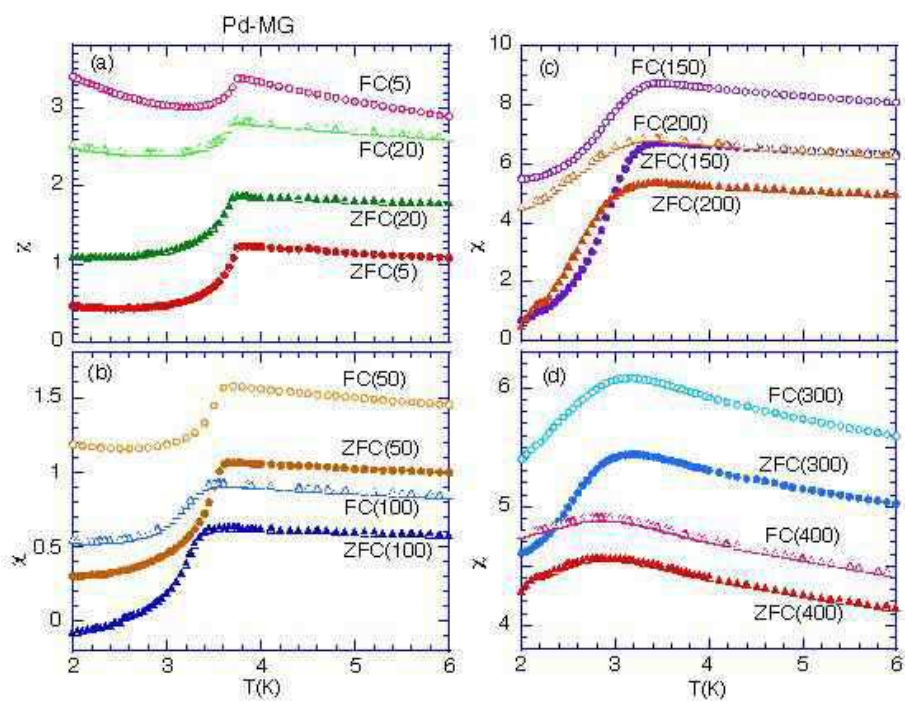}%
\end{center}
\caption{\label{fig:eight}(a) - (d) $T$ dependence of $\chi_{ZFC}$ and
$\chi_{FC}$ for Pd-MG at various $H$.}
\end{figure}

\begin{figure}
\begin{center}
\includegraphics[width=8.0cm]{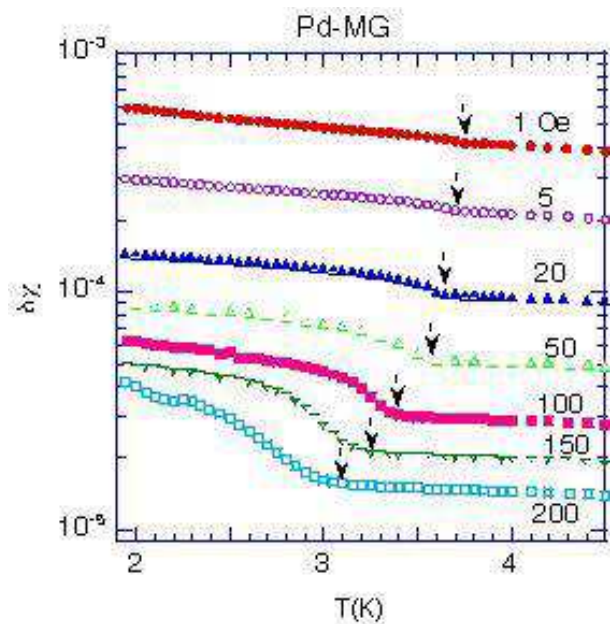}%
\end{center}
\caption{\label{fig:nine}$T$ dependence of the difference $\delta \chi$ (=
$\chi_{FC} - \chi_{ZFC}$) for Pd-MG at various $H$.  The temperature
($T_{\delta}$) denoted by arrow is a characteristic temperature at which
$\delta \chi$ starts to increase drastically with decreasing $T$}
\end{figure}

In the previous paper \cite{Suzuki2000} we show the $T$ dependence of
$\chi_{ZFC}$ (= $M_{ZFC}$/$H$) and $\chi_{FC}$ (= $M_{FC}$/$H$) for Pd-MG
where $H$ = 1 Oe and $1.9 \leq T \leq 298$ K. We find that the deviation of
$\chi_{ZFC}$ from $\chi_{FC}$ starts to appear below 298 K, suggesting a
behavior reminiscent of frustrated spin systems.  In the present work, we
have measured the $T$ dependence of $\chi_{ZFC}$ and $\chi_{FC}$ at low
temperatures below 6 K. Figures \ref{fig:seven} and \ref{fig:eight} (a) -
(d) show the $T$ dependence of $\chi_{ZFC}$ and $\chi_{FC}$ for Pd-MG at
various $H$.  At $H$ = 1 Oe (see figure \ref{fig:seven}), a cusp-like peak
is observed at 3.8 K for both $\chi_{ZFC}$ and $\chi_{FC}$.  A local
minimum is observed around 2.7 K in $\chi_{ZFC}$ and 3.4 K in $\chi_{FC}$. 
As shown in figure \ref{fig:eight}, the cusp-like peak shifts to the
low-$T$ side with increasing $H$, becoming into a broader peak for $H \geq
100$ Oe.  The local minimum of $\chi_{FC}$ at a temperature $T_{min}$ also
shifts to the low-$T$ side with increasing $H$ for $1 \leq H \leq$ 50 Oe. 
The increase of $\chi_{FC}$ below $T_{min}$ indicates the existence of AF
short-range order at low temperatures.  The resultant susceptibility arises
from a competition between a negative diamagnetic susceptibility due to the
Meissner effect and a positive AF susceptibility.  Figure 9 shows the $T$
dependence of the difference $\delta \chi$ defined by $\delta \chi =
\chi_{FC} - \chi_{ZFC}$.  The difference $\delta \chi$ provides a measure
for the irreversible effect of magnetization.  The difference $\delta \chi$
is positive at least below 60 K for $1 \leq H \leq 500$ Oe.  The growth of
the AF spin correlation length is greatly limited by the disordered nature
of nanographites, forming AF short-range order.  Because of the frustrated
nature, the system magnetically behaves like SG's.  Note that $\delta \chi$
starts to increase drastically with decreasing $T$ below a characteristic
temperature $T_{\delta}$, which is nearly equal to the peak temperature
($T_{1}(H)$) of $\chi^{\prime}$ vs $T$ (see section \ref{disA} for the
definition of $T_{1}(H)$).

\subsection{\label{resultE}Aging effect}

\begin{figure}
\begin{center}
\includegraphics[width=8.0cm]{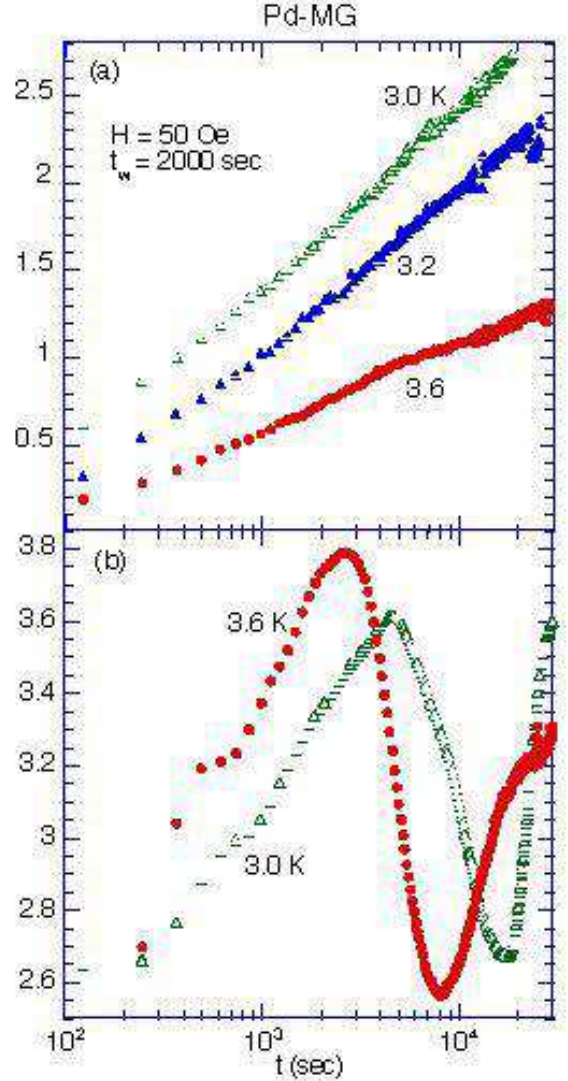}%
\end{center}
\caption{\label{fig:ten}(a) $t$ dependence of $m_{ZFC}$ [$= \Delta
M_{ZFC}(t,T)/ M_{ZFC}(t=0,T)$] at $T$ = 3.0, 3.2, and 3.6 K in the presence
of $H$ = 50 Oe, where $\Delta M_{ZFC}(t,T) = M_{ZFC}(t,T) -
M_{ZFC}(t=0,T)$.  The sample was cooled from 50 K to $T$ at $H$ = 0 and was
kept at $T$ for a wait time $t$$_{w}$ ($= 2.0 \times 10^{3}$ sec) before
$H$ is switched on at $t$ = 0.  (b) $t$ dependence of the relaxation rate
$S(t)$ ($= (1/H)$d$M_{ZFC}/$d$\ln t)$ at $T$ = 3.0 and 3.6 K, which is
estimated from figure \ref{fig:ten}(a).}
\end{figure}

The relaxation of the ZFC magnetization was measured.  The system was
rapidly cooled from 50 K to $T$ (= 3.0, 3.2, and 3.6 K) in the absence of
$H$.  After the isothermal aging was carried out at $T$ and $H$ = 0 for a
wait time $t_{w}$ $= 2.0 \times 10^{3}$ sec, the field $H$ (= 50 Oe) is
switched on at $t$ = 0.  Figure \ref{fig:ten}(a) shows the time ($t$)
dependence of $m_{ZFC}(t,T)$ [$= \Delta M_{ZFC}(t,T)/M_{ZFC}(t=0,T)$],
where $\Delta M_{ZFC}(t,T) = M_{ZFC}(t,T) - M_{ZFC}(t=0,T)$.  The
magnetization $m_{ZFC}(t,T)$ at $T$ = 3.0, 3.2 K is almost proportional to
$\ln t$ for $t < t_{w}$ and it deviates from the $\ln t$ dependence for $t
> t_{w}$.  The $t$ dependence of $m_{ZFC}(t,T)$ at $T$ = 3.6 K is rather
different from that at $T$ = 3.0 and 3.2 K. Figure \ref{fig:ten}(b) shows
the relaxation rate $S(t)$ defined by $S(t) = (1/H)$d$M_{ZFC}(t)/$d$\ln t$
at $T$ = 3.0 and 3.6 K, which is calculated from the data shown in figure
\ref{fig:ten}(a).  The relaxation rate $S(t)$ exhibits a broad peak at a
characteristic time $t_{cr}$.  The time $t_{cr}$ linearly decreases with
increasing $T$: $t_{cr}$ is longer than $t_{w}$ for $3.0 \leq T \leq 3.6$ K
and is assumed to be equal to $t_{w}$ at $T$ = 3.78 K. The width of the
peak in $S(t)$ at 3.0 K is broader than that at 3.6 K, indicating a broader
distribution of relaxation times as $T$ is lowered.  This phenomenon is
called aging and has been observed in SG's
\cite{Nordblad1986,Granberg1988}.  Similar $T$ dependence of $t_{cr}$ has
been observed in the 3D Ising SG system
Cu$_{0.5}$Co$_{0.5}$Cl$_{2}$-FeCl$_{3}$ graphite bi-intercalation compound
(GBIC) \cite{Suzuki2003}, where $t_{cr}$ almost linearly decreases with
increasing $T$ and becomes equal to $t_{w}$ at the spin freezing
temperature $T_{SG}$.  As far as we know, the aging behavior of $M_{ZFC}$
has never been observed in usual superconductors.  However, there is only
one exception for this.  The aging, rejuvenation, and memory effects have
been reported in a melt-cast Bi$_{2}$Sr$_{2}$CaCu$_{2}$O$_{8}$ sample
displaying a phenomenon called paramagnetic Meissner (PME) effect
($M_{FC} > 0$ and $M_{ZFC} < 0$ below $T_{c}$)
\cite{Papadopoulou1998,Papadopoulou2001}.  In this case, the competing
interactions between current loops (magnetic moments) give rise to a glassy
low-temperature state, where the aging behavior is similar to that observed
in SG's.  The aging behavior observed in Pd-MG is magnetic in origin,
suggesting the existence of the SG-like behavior below $T_{c}$.

\subsection{\label{resultF}DC magnetic susceptibility at high fields}

\begin{figure}
\begin{center}
\includegraphics[width=8.0cm]{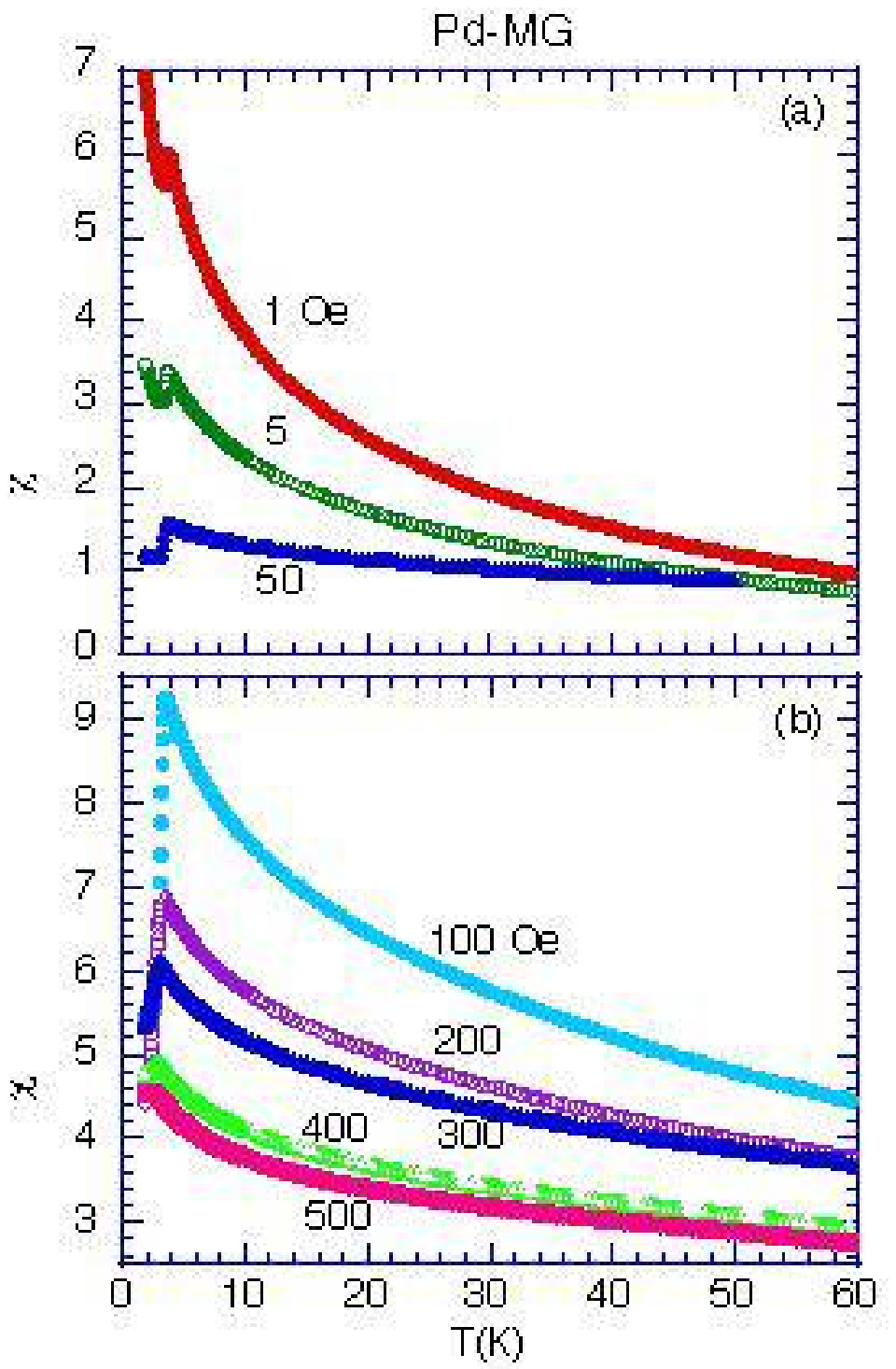}%
\end{center}
\caption{\label{fig:eleven}(a) and (b) $T$ dependence of $\chi_{FC}$ for
Pd-MG at various $H$.}
\end{figure}

\begin{figure}
\begin{center}
\includegraphics[width=8.0cm]{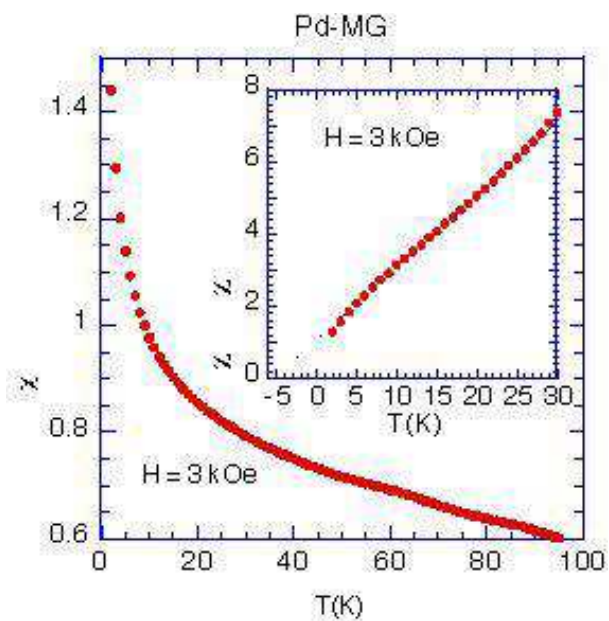}%
\end{center}
\caption{\label{fig:twelve}$T$ dependence of $\chi_{FC}$ for Pd-MG at $H$
= 3 kOe.  The inset shows the reciprocal susceptibility ($\chi_{FC} -
\chi_{0})^{-1}$.}
\end{figure}

\begin{figure}
\begin{center}
\includegraphics[width=8.0cm]{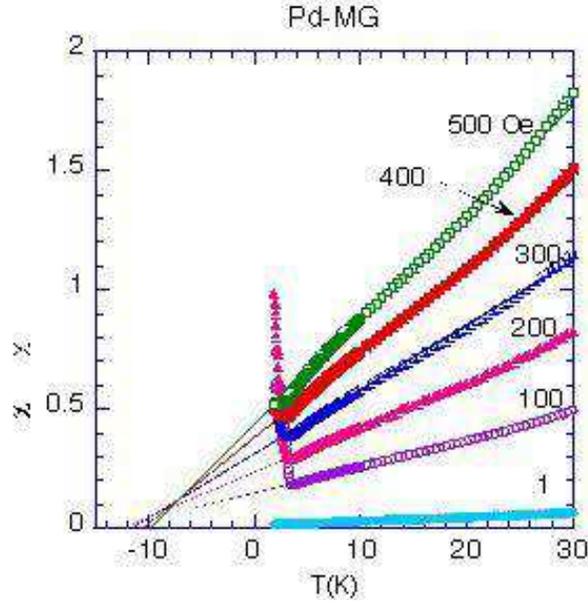}%
\end{center}
\caption{\label{fig:thirteen}$T$ dependence of the reciprocal
susceptibility ($\chi_{FC} - \chi_{0})^{-1}$ for Pd-MG at various $H$,
where $\chi_{0}$ is a $T$-independent susceptibility determined from the
least squares fit of the data to the Curie-Weiss law (\ref{eq:one}) for
each $H$.  The solid straight lines denote the fitting curves, where
$C_{g}$, $\Theta$, and $\chi_{0}$ are listed in the table \ref{table:one}.}
\end{figure}

\begin{table}
\caption{\label{table:one}Curie-Weiss constant $C_{g}$, Curie-Weiss
temperature $\Theta$, and $T$- independent susceptibility $\chi_{0}$
determined from the least-squares fit of the data of $\chi_{FC}$ vs $T$ at
various $H$ in the temperature range between 5 and 30 K.}
\begin{indented}
\item[]\begin{tabular}{@{}llll}
\br
$H$ (Oe) & $C_{g}$ (10$^{-3}$ emu K/g) & $\Theta$ (K) & $\chi_{0}$ 
(10$^{-5}$ emu/g)\\
\mr
1 & 5.81 & -6.9 & 4.02\\
5 & 3.85 & -9.4 & 3.94\\
100 & 0.90 & -13.1 & 3.72\\
200 & 0.53 & -12.4 & 3.40\\
300 & 0.37 & -11.5 & 3.47\\
400 & 0.28 & -10.6 & 2.76\\
500 & 0.23 & -9.8 & 2.63\\
3000 & 0.05 & -5.4 & 0.66\\
\br
\end{tabular}
\end{indented}
\end{table}

Figures \ref{fig:eleven} (a) and (b) show the $T$ dependence of $\chi_{FC}$
of Pd-MG for $H$ = 1 - 500 Oe.  Figure \ref{fig:twelve} shows the $T$
dependence of $\chi_{FC}$ at $H$ = 3 kOe.  The susceptibility obeys the
Curie-Weiss law only at low temperatures ($5 \leq T \leq 30$ K).  
The least-squares fit of the data $\chi_{FC}$ vs $T$ for each $H$ to
\begin{equation}
\chi_{FC} = \chi_{0} + C_{g}/(T-\Theta),
\label{eq:one}
\end{equation}
yields the Curie-Weiss constant $C_{g}$, the Curie-Weiss temperature
$\Theta$, and $T$-independent susceptibility $\chi_{0}$, which are listed
in Table I. The values of $C_{g}$, $\Theta$, and $\chi_{0}$ are different
for different $H$.  The inset of figure \ref{fig:twelve}, and figure
\ref{fig:thirteen} show the $T$ dependence of the corresponding reciprocal
susceptibility ($\chi_{FC} - \chi_{0})^{-1}$ for $H$ = 3 kOe, and $1 \leq H
\leq 500$ Oe, respectively.  We find that $\Theta$ is negative for any $H$,
suggesting the AF nature of the system.  The inclusion of the data for $T
\geq$ 30 K gives rise to a deviation of the data from (\ref{eq:one}).  This
is in contrast to the paramagnetic susceptibility due to magnetic
impurities, where the agreement of the data with (\ref{eq:one}) becomes
better as the data at higher $T$ are included.  Since $\Theta$ is close to
zero, hereafter we call this behavior the Curie-like behavior rather than
Curie-Weiss like behavior.  Such a Curie-like behavior is due to the
localized conduction electrons near zigzag edge sites of nanographites. 
Each electron has the effective magnetic moment $P_{eff} =
g[S(S+1)]^{1/2}$, where the Land\'{e} $g$ factor $g$ = 2 and spin $S$ = 1/2
of the conduction electron.  Then the value of $N_{g}$, the number of spins
of localized conduction electrons (per gram), can be estimated as $N_{g} =
(3k_{B}C_{g}/\mu_{B}^{2}P_{eff}^{2}) = 8.0 \times 10^{19}$ per gram, where
we use $P_{eff} = \sqrt{3}$ and $C_{g} = 4.98 \times 10^{-5}$ emu K/g for
$H$ = 3 kOe.  This value of $N_{g}$ is a little larger than that reported
by Shibayama et al.  \cite{Shibayama2000} for activated carbon fibers (ACF)
composed of a disorder network of nanographites [$N_{g} = (0.39 - 4.2)
\times 10^{19}$/g].

\begin{figure}
\begin{center}
\includegraphics[width=8.0cm]{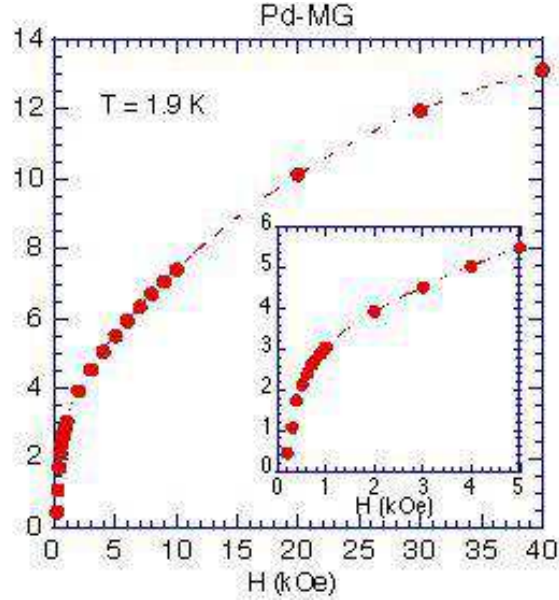}%
\end{center}
\caption{\label{fig:fourteen}$H$ dependence of $M_{ZFC}$ at 1.9 K for
Pd-MG. After the sample was cooled from 298 to 1.9 K at $H$ = 0, the
measurement was carried out with increasing $H$ from 0 to 40 kOe.  The
inset shows the detail of $M_{ZFC}$ vs $H$ at low $H$.  The solid lines are
guides to the eyes.}
\end{figure}

Figure \ref{fig:fourteen} shows the $H$ dependence of $M_{ZFC}$ at 1.9 K
for $0 \leq H \leq 40$ kOe.  The magnetization $M_{ZFC}$ exhibits strong
nonlinear $H$ dependence, and it reaches 0.13 emu/g at $H$ = 40 kOe, which
does not correspond to the saturation magnetization $M_{s}$.  It is
predicted that the ratio of the saturation magnetization $M_{s}$ ($=
N_{g}\mu_{B}gS$) to $C_{g}$ is given by $3k_{B}/[\mu_{B}g(S+1)]$, which is
independent of $N_{g}$.  When $g$ = 2 and $S$ = 1/2 for the conduction
electron, the saturation magnetization $M_{s}$ is estimated as 0.74 emu/g. 
This value of $M_{s}$ is much larger than that of $M_{ZFC}$ at $H$ = 40
kOe.  The reason for such a large $M_{s}$ is not clear.

\section{\label{dis}DISCUSSION}
\subsection{\label{disA}Possibility of a quasi-2D superconductivity in Pd sheets}

\begin{figure}
\begin{center}
\includegraphics[width=8.0cm]{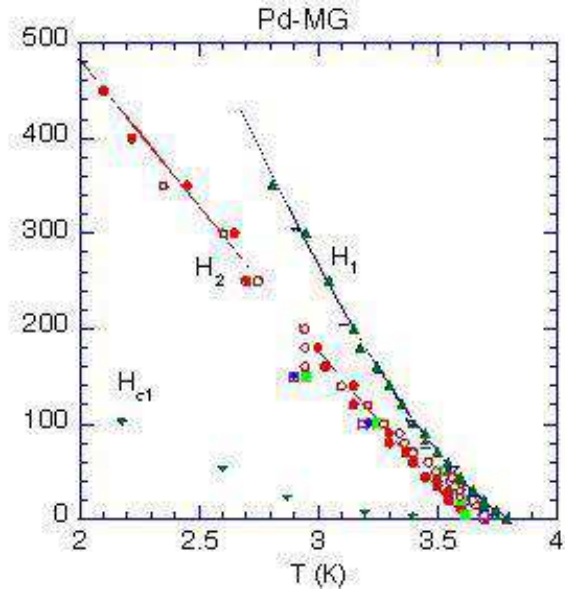}%
\end{center}
\caption{\label{fig:fifteen}$H$-$T$ diagram for Pd-MG, where the peak
temperatures of $\chi^{\prime}$ vs $T$ (closed triangle),
d$\chi^{\prime}$/d$T$ vs $T$ (closed circle), $\chi^{\prime\prime}$ vs $T$
(open circle), d$\chi_{ZFC}$/d$T$ vs $T$ (open square), and
d$\chi_{FC}$/d$T$ vs $T$ (closed diamond), the local-minimum temperatures
of d$\delta \chi$/d$T$ vs $T$ (closed square) and $\chi_{FC}$ vs $T$
(closed triangle down), and $T_{\delta}$ for $\delta \chi$ vs $T$ (open
triangle down), are plotted as a function of $H$.  The superconducting
transition temperature $T_{c}(H)$ coincide with the peak temperatures
($T_{2}(H)$) of d$\chi^{\prime}$/d$T$ vs $T$ and $\chi^{\prime\prime}$ vs
$T$.  The solid lines are least-squares fitting curves.  See the text in
detail.}
\end{figure}

Our results obtained above are summarized in the $H$-$T$ diagram of Pd-MG.
In figure \ref{fig:fifteen} we make a plot of the peak temperatures of
$\chi^{\prime}$ vs $T$, d$\chi^{\prime}$/d$T$ vs $T$, $\chi^{\prime\prime}$
vs $T$, d$\chi_{ZFC}$/d$T$ vs $T$, and d$\chi_{FC}$/d$T$ vs $T$, the
local-minimum temperatures of d$\delta \chi$/d$T$ vs $T$ and $\chi_{FC}$ vs
$T$, and the characteristic temperature $T_{\delta}$ for $\delta \chi$ vs
T, as a function of $H$.  The data are classed in three groups: (i) the
upper line for $\chi^{\prime}$ vs $T$ and $\delta \chi$ vs $T$ (denoted as
the line $H_{1}(T)$ or the line $T_{1}(H)$), (ii) the intermediate line for
d$\chi^{\prime}$/d$T$ vs $T$, $\chi^{\prime\prime}$ vs $T$,
d$\chi_{ZFC}$/d$T$ vs $T$, d$\chi_{FC}$/d$T$ vs $T$, and d$\delta
\chi$/d$T$ vs $T$ (denoted as the line $H_{2}(T)$ or the line $T_{2}(H)$),
and (iii) the lower line for $\chi_{FC}$ vs $T$ which may correspond to the
lower critical line $H_{c1}(T)$ or the line $T_{c1}(H)$.  Here we assume
that the $T$ dependence of the line $H_{j}$ ($j$ = 1 and 2) may be
described by a power law form given by
\begin{equation}
H = H_{j}^{*}(1-T/T_{j})^{\alpha_{j}},
\label{eq:two}
\end{equation}
where $\alpha_{j}$ is an exponent, and $H_{j}^{*}$ and $T_{j}$ are the
characteristic field and critical temperature, respectively.  The
least-squares fit of the data of $H_{1}$ vs $T$ for $40 \leq H \leq 350$ Oe
to (\ref{eq:two}) with $j$ = 1 yields the parameters $\alpha_{1} = 1.43 \pm
0.05$, $T_{1} = 3.82 \pm 0.04$ K, and $H_{1}^{*} = 2.40 \pm 0.05$ kOe.  The
exponent $\alpha_{1}$ is close to that (= 1.50) for the irreversibility
line [the so-called Almeida-Thouless (AT) line] \cite{Almeida1978}.  In
fact, as shown in figure 9, the difference $\delta \chi$ starts to increase
with decreasing $T$ below $T_{1}(H)$.  The least-squares fit of the data of
$H_{2}$ vs $T$ for $0 \leq H \leq 450$ Oe to (\ref{eq:two}) with $j$ = 2
yields the parameters $\alpha_{2} = 1.06 \pm 0.07$, $T_{2} = 3.63 \pm 0.04$
K, and $H_{2}^{*} = 1.12 \pm 0.04$ kOe.  Note that the upper critical field
$H_{c2}$ of the type-II superconductor is defined by $H_{c2} =
\Phi_{0}/(2\pi \xi^{2})$, where $\Phi_{0}$ ($= 2.0678 \times 10^{-7}$
Gauss cm$^{2}$) is a fluxoid and $\xi$ is the coherence length
\cite{Ketterson1998}.  Since $\xi$ is proportional to $(1 -
T/T_{c})^{-1/2}$ according to the Ginzburg-Landau theory (mean-field
theory), it is predicted that $H_{c2}$ is proportional to $(1 - T/T_{c})$
\cite{Ketterson1998}.  The exponent $\alpha_{2}$ is slightly larger than
the mean-field value (= 1).  Thus it is reasonable to assume that $T_{2}$
coincides with $T_{c}$.  In fact, $T_{2}(H)$ corresponds to a peak temperature
of d$\chi^{\prime}$/d$T$ vs $T$, which is defined as a critical temperature
for usual superconductors.  The upper critical field $H_{c2}(T=0$ K) 
is
roughly estimated from the slope of the line $H_{2}(T)$ at $T = T_{c}$:
(-d$H_{2}$/d$T)_{T_{c}} = 180 \pm 20$ Oe/K. Using the relation $H_{c2}(0)
\approx -0.69T_{c}($d$H_{2}$/d$T)_{T_{c}}$ \cite{Werthamer1966}, the
extrapolated $H_{c2}(0)$ is estimated as 450 $\pm$ 50 Oe.  The coherence
length $\xi(0)$ is estimated as $\xi(0) = 850 \pm 10 \AA$, which is larger
than the average size of Pd nanoparticles ($= 530 \pm 340) \AA$.

It is well known that the $H$-$T$ diagram of an ideal quasi-2D
superconductors such high $T_{c}$ superconductors consists of a vortex
lattice (Abrikosov lattice) phase, vortex glass phase, and a vortex liquid
phase \cite{Blatter1994,Gammel1998}.  The three lines ($H_{al}$, $H_{ag}$,
and $H_{gl}$ merge into a multicritical point around $T^{*}$ and $H^{*}$,
where the line $H_{al}$ is the boundary between the vortex lattice and the
vortex liquid phase, the line $H_{ag}$ is the boundary between the vortex
lattice phase and the vortex glass phase, and the line $H_{gl}$ is the
boundary between the vortex glass phase and vortex liquid phase.  Our
$H$-$T$ diagram of Pd-MG is compared with that of typical quasi-2D
superconductors.  It seems that the lines $H_{1}$ and $H_{2}$ correspond to
the lines $H_{gl}$ and $H_{ag}$, respectively.  The line $H_{1}$ is the
irreversibility line similar to the line $H_{gl}$.  While the lines
$H_{gl}$ and $H_{ag}$ merge at a multicritical point, the lines $H_{1}$ and
$H_{2}$ do not merge even at $H$ = 0.  The absence of the multicritical
point and the line $H_{al}$, suggests that the superconductivity in Pd-MG is
quasi-2D.

So far we assume that the superconductivity occurs mainly in Pd sheets and
that the AF short-range order occurs in nanographites (denoted by model I). 
There may be another possibility that the superconductivity occurs in
nanographites and that the AF short-range order occurs in the Pd sheets
(denoted by model II).  This possibility can be ruled out for the following
reason.  Experimentally, Kopelevich et al.  \cite{Kopelevich2000} have
reported the superconducting-like magnetization hysteresis loops in highly
oriented pyrolytic graphite (HOPG).  Theoretically, Gonz\'{a}lez et al. 
\cite{Gonzalez2001} have predicted that a topological disorder in graphene
sheets can trigger the instabilities for a possible $p$-wave
superconductivity.  Pd-MG exhibits the type-II superconductivity.  There is a
relationship between the upper critical field $H_{c2}$ and $T_{c}$: $H_{c2}
= 124 T_{c}$ in the units of K for $T_{c}$ and Oe for $H_{c2}$.  This
relation is in good agreement with an empirical law ($H_{c2} \approx 200
T_{c}$) derived from the BCS (Bardeen-Cooper-Schrieffer) theory
\cite{Ketterson1998}, suggesting that the conventional superconductivity
occurs in Pd-MG. In this sense, the model I is preferable to the model II,
in spite of the lack in our knowledge in the possible $p$-wave superconductivity in
graphene sheets.

\subsection{\label{disB}AF short-range order and SG-like behavior in graphene 
sheets}
The nature of the AF short-range order is discussed.  The existence of AF
short-range order has been confirmed from the following results.  (i) The
susceptibility $\chi_{FC}$ obeys a Curie-Weiss law with a negative
$\Theta$.  The interactions between localized magnetic moments is
antiferromagnetic.  (ii) The irreversibility between $\chi_{ZFC}$ and
$\chi_{FC}$ occurs well above $T_{c}$.  The growth of spin order is greatly
suppressed by the disordered nature of nanographites, forming the AF
short-range order.  Because of the frustrated nature of AF interaction
between nanographites, no long-range order can develop at any finite
temperatures.  We have found several results supporting that the AF
short-range order below $T_{c}$ exhibits a SG-like behavior.  (i) For an
usual superconducting phase, $M_{TR}$ is larger than $M_{FC}$, and $M_{IR}$
is larger than $M_{IR}$ below $T_{c}(H)$.  For Pd-MG, $M_{TR}$ is larger than
$M_{FC}$, and $M_{IR}$ is larger than $M_{IR}$ for $T < T_{\alpha} <
T_{2}(H)$, while $M_{TR}$ is smaller than $M_{FC}$, and $M_{IR}$ is smaller
than $M_{IR}$ for $T_{\alpha} < T < T_{2}(H)$.
The latter is a feature common to the SG systems having spin frustration
effect.  (ii) The aging dynamics of $M_{ZFC}(t)$ is observed below $T_{c}$,
corresponding to a slow evolution of non-equilibrium spin configurations
towards equilibrium ones.  The relaxation rate $S(t) =
(1/H)$d$M_{ZFC}$/d$\ln t$ shows a broad peak at $t_{cr}$ which is longer
than $t_{w}$.  Such an aging behavior is usually observed in SG systems. 
When the SG system is quenched from a high temperature above 
$T_{SG}$ to a low temperature $T$ below $T_{SG}$,
the initial state is not thermodynamically stable and relaxes to more
stable state.  The aging behaviors depend strongly on their thermal history
within the SG phase.  Note that a SG-like behavior has been observed in
$\chi_{FC}$ of ACF which is composed of disordered network of nanographites
\cite{Shibayama2000}.  The susceptibility $\chi_{FC}$ of ACF shows a cusp
around 4 - 7 K. This behavior is understood in terms of spin frustration
effect arising from random strengths of the AF interactions between
nanographites.  As is discussed in section \ref{disA}, the line $H_{1}(T)$
is assumed to be the irreversibility line for the magnetization in the
superconductivity (the line $H_{gl}$).  However, there is some 
possibility that
the line $H_{1}(T)$ may correspond to the AT line for the SG-like
behavior.

Harigaya \cite{Harigaya2001a,Harigaya2002,Harigaya2001b} has theoretically
predicted that the magnetism in nanographites with zigzag edge sites
depends on the stacking sequence of nanographites.  For the $A$-$B$
stacking, there is no interlayer interaction $J_{1}$ at the edge site,
where $J_{1}$ is the strength of the weak hopping interaction between
neighboring layers.  This gives rise to the finite magnetic moment.  The AF
spin alignment is favorable for strong on-site repulsion $U$.  The local
magnetic moments tend to exist at the edge sites in each layer due to the
large amplitude of the wavefunctions at these sites.  The number of up-spin
electrons is larger than that of down-spin electrons in the first layer.  The
number of down-spin electrons is larger than that of up-spin electrons in
the second layer, and so on.  The system can be described by the AF
Heisenberg model with the exchange interaction $J^{\prime}$ ($=
2J_{1}^{2}/U$).  For the $A$-$A$ type stacking, on the other hand, the
magnetic moment per layer does not appear due to the interlayer
interaction.  The up- and down-spin electrons are not magnetically
polarized in each layer.  Although there is no detailed structural study on the
stacking sequence in Pd-MG, the AF order in Pd-MG may suggest that the
$A$-$B$ stacking is dominant compared to the $A$-$A$ stacking.  The spin
correlation length along the $c$ axis is considered to be on the same order
of the nearest neighbor interlayer distance between nanographites.  Since
the Pd sheet is sandwiched between graphene sheets, there is no net
molecular field on Pd sheet from graphene sheets having the $A$-$B$
stacking along the $c$ axis.  This may lead to the coexistence of
superconductivity and AF short-range order.

\subsection{\label{disC}Effect of multilayered Pd sheets}
In an ideal Pd-MG, there is one Pd sheet (monolayer) between adjacent
graphene sheets.  In reality, there are either monolayer or multilayers (2
- 4 layers) between graphene sheets.  Multilayered Pd nanoparticles would
generate internal stress inside the graphite lattice, leading to the break
up of adjacent graphene sheets into nanographites.  How does the spin
fluctuation vary with the number of Pd layers in a quasi 2D-like 
Pd systems consisting of stacked Pd layers?  Bouarab et
al. \cite{Bouarab1990} have predicted (i) no magnetic moment for $n$ = 1, (ii)
ferromagnetic moment for $n$ = 2 - 5, and (iii) no magnetic moment for $n >
5$, where $n$ is the number of Pd layers.  The value of $N(E_{F})$ for $n$
= 1 is smaller than that in the bulk Pd.  The 2D effect increases the DOS
in the middle of the energy band, whereas it decreases the DOS in the
higher energy side where $E_{F}$ exists.  Because the Stoner criterion
$J_{s}N(E_{F}) > 1$ is not satisfied, the Pd monolayer is non-magnetic,
where $J_{s}$ is an exchange parameter.  The peak of the DOS, which is much
below $E_{F}$ for $n$ = 1, moves towards $E_{F}$ for higher $n$.  This
prediction suggests that the Pd monolayer may be a superconductor because
of the suppression of spin fluctuations.  On the other hand, Pd layers with
$n$ = 2 - 5 may be a ferromagnet.  In Pd-MG, at present it is not clear
what is the minimum number of Pd layers ($n_{c}$) required for the
ferromagnetic state.  The value of $n_{c}$ is dependent on the size of Pd
nanoparticles.  However, it is reasonable to conclude that Pd layers in
Pd-MG is ferromagnetic for $n \geq n_{c}$ corresponding to the sample with
the long-reaction time and is superconducting for $n < n_{c}$ corresponding
to the sample with the short-reaction time \cite{Suzuki2000}.

\section{CONCLUSION}
Pd-MG undergoes a superconducting transition at $T_{c}$ (= 3.63 K).  A
quasi-2D superconductivity occurs in Pd sheets.  The AF short-range order
appears well above $T_{c}$ in graphene sheets.  The growth of the AF
short-range order is limited by the disordered nature of nanographites. 
Both the aging behavior of $M_{ZFC}(t)$ and the $T$ dependence of
$M_{ZFC}$, $M_{FC}$, $M_{IR}$, and $M_{TR}$ below $T_{c}$ suggest the
existence of a SG like behavior in nanographites.  Further studies are
required to understand the possible interplay between the superconductivity
and SG-like behavior, including the nature of the irreversibility line
$H_{1}$($T$).

\ack
The authors would like to thank K. Harigaya for valuable discussions on the
antiferromagnetism in nanographites.  The work at Binghamton was supported
by the Research Foundation of SUNY-Binghamton (contract number 240-9522A). 
The work at Osaka (J.W.) was supported by the Ministry of Education,
Science, Sports and Culture, Japan [the grant for young scientists (No. 
70314375)] and by Kansai Invention Center, Kyoto, Japan.

\end{document}